\documentclass[twocolumn, amssymb, amsmath, aps, prb, showpacs, 10pt]{revtex4-1}
\usepackage[dvips]{graphicx}
\usepackage[dvips]{color}
\begin{document}
\title{Hydrostatic pressure tuned magneto-structural transition and occurrence of pressure induced exchange bias effect in Mn$_{0.85}$Fe$_{0.15}$NiGe alloy}
\author{P. Dutta$^1$}
\author{S. Pramanick$^2$}
\author{D. Das$^1$}
\author{S. Chatterjee$^1$}
\email{souvik@alpha.iuc.res.in}
\affiliation{$^1$UGC-DAE Consortium for Scientific Research, Kolkata Centre, Sector III, LB-8, Salt Lake, Kolkata 700 098, India}
\affiliation{$^2$Department of Solid State Physics, Indian Association for the Cultivation of Science, 2 A \& B Raja S. C. Mullick Road, Jadavpur, Kolkata 700 032, India}
\begin{abstract}
Magnetic and magneto-functional behavior of a Fe-doped MnNiGe alloy with nominal composition Mn$_{0.85}$Fe$_{0.15}$NiGe have been investigated in ambient as well as in high pressure condition. The alloy undergoes first order martensitic phase transition (MPT) around 200 K and also shows large conventional magnetocaloric effect (MCE) ($\Delta S$ $\sim$ -21 J/kg-K for magnetic field ($H$) changing from 0-50 kOe) around the transition in ambient condition. Application of external hydrostatic pressure ($P$) results a shift in MPT towards the lower temperature and a clear decrease in the saturation moment of the alloy at 5 K. The peak value of MCE is also found to decrease with increasing external $P$ ($\sim$ 18 J/kg-K decrease in $\Delta S$ has been observed for $P$ = 12.5 kbar). The most interesting observation is the occurance of exchange bias effect (EBE) on application of external $P$. The competing ferromagnetic and antiferromagnetic interaction in presence of external $P$ plays the pivotal role towards the observation of $P$ induced EBE.
\end{abstract}
\maketitle

\section{Introduction}

Magnetic equiatomic alloys (MEAs) of general formula MM$^\prime$X (M, M$^\prime$ = transition metals, X = Si, Ge, Sn etc.) and their derivatives have attracted renewed attentions for the observation of fascinating magneto-functional and physical properties such as, large magneticaloric effect (MCE), exchange bias effect (EBE), spin glass (SG) like ground state etc.~\cite{zhang-apl,Koyama-fsma,zhang-jpd,pd-epl,ali-apl,Trung-apl1,pd-jmmm,liu-nc,Caron-prb1,Dincer-jalcom1,Trung-mncoge,Wang-mncoge} Among various members of MEAs, MnNiGe is one of the potential candidates which undergoes a first-order diffusionless structural phase transition, known as  martensitic phase transition (MPT), at 470 K during cooling and orders spiral antiferromagnetically below 346 K.~\cite{zhang-jpd,zhang-apl,pd-epl,liu-nc} Physical and chemical pressures are the two most influencing parameters that can affect the physical properties of these materials simply by modifying structural parameters like lattice volume, bond angle etc.~\cite{Anzai-pressure,Niziol-pressure,liu-nc} Among these two, chemical pressure approach (different doping studies) is much popular among the researchers due to easy sample preparation and measurement options. However, the purest form of perturbation can only be given by the physical pressure approach. Since MEA's discovery, several doping studies (chemical pressure effect) have been performed to increase the magneto-functional properties by reducing the structural transition temperature below its magnetic transition.~\cite{zhang-jpd,pd-epl,ali-apl,Trung-apl1,pd-jmmm,liu-nc} But very little efforts have been made to tune the magneto-functional properties by applying physical pressure on the MnNiGe system.~\cite{Anzai-pressure,samanta-pressure,Niziol-pressure}  Recent investigation on the Fe-doped MnNiGe alloy (Fe doping both in Mn and Ni sites of the alloy) by Liu {\it et al.} indicates the presence of ferromagnetic (FM) and SG-like ground state in the system.~\cite{liu-nc} Fe doping at the Mn site of the MnNiGe layered compound not only breaks the spiral antiferromagnetic (AFM) Mn-Ge-Mn inter-layer interaction but also induces FM interaction between Mn atoms of the same layer by reducing lattice volume of the alloy.~\cite{liu-nc}  In addition, a reasonable decrease in MPT temperature of the alloy has been observed with Fe-doping. Initial increase in Fe-concentration (up to 16\% doping)  results enhancement of FM strength and hence the overall magnetic moment of the alloy. FM strength again starts to decrease with further increase in Fe concentration and a SG like ground state was observed.~\cite{liu-nc,pd-epl} The direct exchange nature of the Mn-Mn intra-layer interaction plays the pivotal role towards the observation of non-monotonic effect of Fe-doping. In the present work, our aim is to address the effect of clean perturbation by applying external hydrostatic pressure on the structural and magnetic transition of a Fe-doped MnNiGe alloy with nominal composition Mn$_{0.85}$Fe$_{0.15}$NiGe. 

\begin{figure}[t]
\centering
\includegraphics[width = 8.5 cm]{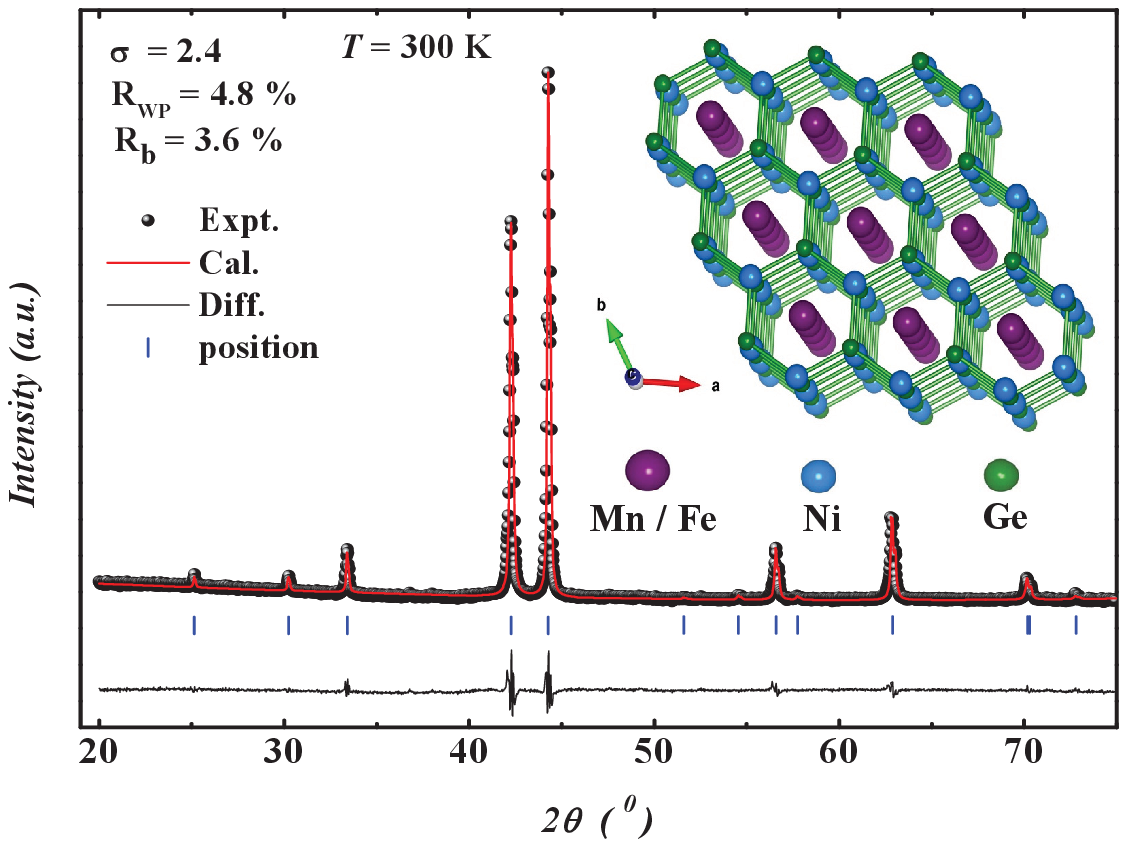}
\caption{(Color online)  The observed (circle symbol) and calculated (solid line, fitted by peak profile Rietveld refinement) X-ray powder diffraction pattern at room temperature of Mn$_{0.85}$Fe$_{0.15}$NiGe compound with difference pattern (residue) at the bottom. Inset shows the structure of the alloy at room temperature.} 
\label{fig1}
\end{figure}

\section{Experimental Details}

The polycrystalline sample of nominal composition Mn$_{0.85}$Fe$_{0.15}$NiGe was prepared by  argon arc-melting the constituent elements (purity $>$ 99.9\%). To make the alloy homogeneous, the ingot was turned over and re-melted for four times. Further the ingot was sealed in an evacuated quartz tube and annealed at 800$^{\circ}$C for 100 h followed by a rapid quenching in ice water. The sample was characterized by room temperature x-ray powder diffraction (XRD) measurement recorded in a Bruker AXS diffractometer (D8-Advance) using Cu K$_{\alpha}$ radiation. The collected powder diffraction pattern was used for the Rietveld refinement analysis by MAUD software to obtain different structural parameters.~\cite{maud-jap} Rietveld refinement along with the XRD patterns of the studied sample are shown in the main panel of fig.~\ref{fig1}. XRD analysis confirms that at room temperature, the sample crystalizes with hexagonal Ni$_2$In-type structure (space group $p6_3/mmc$). No noticeable contribution of any impurities has been detected in this pattern. The lattice parameters of the hexagonal phase are found to be $a$ = 4.081 \AA, and $c$ = 5.327 \AA. The important fitted parameters are $\sigma$ = 2.4, $R_{wp}$ = 4.8\%,  and $R_{b}$ = 3.6 \%. The inset of fig.~\ref{fig1} shows the layered structure of hexagonal Mn$_{0.85}$Fe$_{0.15}$NiGe alloy at room temperature obtained by using VESTA software package. The Mn/Fe, Ni and Ge atoms occupy 2a, 2d and 2c sites respectively.~\cite{bazela-mnnige,liu-nc} The dc magnetization ($M$) was measured using a Quantum Design SQUID magnetometer (MPMS 7, Evercool model). High pressure measurements were performed using a MPMS high pressure capsule cell 8.5. Daphne was used as pressure transferring medium.

\section{Results and Discussion}
\begin{figure}[t]
\centering
\includegraphics[width = 8.5 cm]{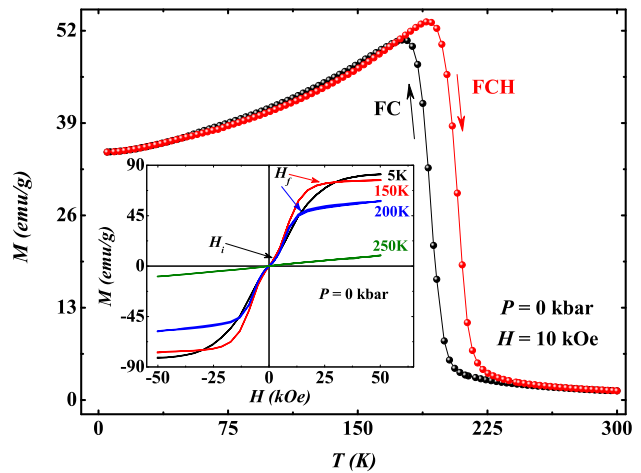}
\caption{(Color online) Temperature ($T$) dependence of magnetization ($M$) in presence of 10 kOe of applied magnetic field ($H$) in the field-cooled (FC) and field-cooled heating (FCH) protocols. Inset depicts isothermal $M$ as a function of applied $H$ at different constant temperature.} 
\label{fig2}
\end{figure}

Temperature ($T$) variation of dc magnetization, measured at an applied magnetic field ($H$) of 10 kOe, both in field cooling (FC) and field-cooled heating (FCH) protocols are depicted in the main panel of fig.~\ref{fig2}. During cooling, a sharp increase in the $M(T)$ data around 200 K in association with a large thermal hysteresis between FC and FCH data has been observed and consequently signify the occurrence of magneto-structural transition in the alloy. Presence of thermal hysteresis confirms the first order nature of the transition. Further decrease in temperature results a sluggish drop in $M$ value after reaching maximum just below the structural transition. The decreasing nature of $M(T)$ data indicates the presence of a short range AFM phase in the otherwise FM alloy. To shed more light on the magnetic nature of the sample, we recorded isothermal $M$ data as a function of $H$ at selected temperatures (see inset of fig.~\ref{fig2}). All the  isothermal $M(H)$ data were recorded during heating at different constant temperatures in a thermally demagnetized state. Paramagnetic nature of the high temperature austenite phase is clear from the linear $M(H)$ data at 250 K. However, closer view of the $M(H)$ data reveals that, below and around MPT, there is a clear signature of change in slope in the $M(H)$ isotherms around the low field region (marked by $H_i$). The isotherms at 200 K and 150 K, when the sample is well in the martensitic state, are associated with field induced AFM to FM transition, which starts at $H_i$ and found to be completed at $H_f$ (marked in the inset of fig.~\ref{fig2}) and the alloy becomes totally FM. Similar behavior has also been observed at 5 K. Both $H_i$ and $H_f$ are found to increase with decreasing temperature. This type of field induced transition from a spiral AFM to canted FM state has already been reported for MnNiGe alloy.~\cite{bazela-mnnige,Fjellvag-mnnige,liu-nc} Magnetic moment at 50 kOe of applied $H$ is also found to increase with decreasing temperature of measurement.

\begin{figure}[t]
\centering
\includegraphics[width = 8.5 cm]{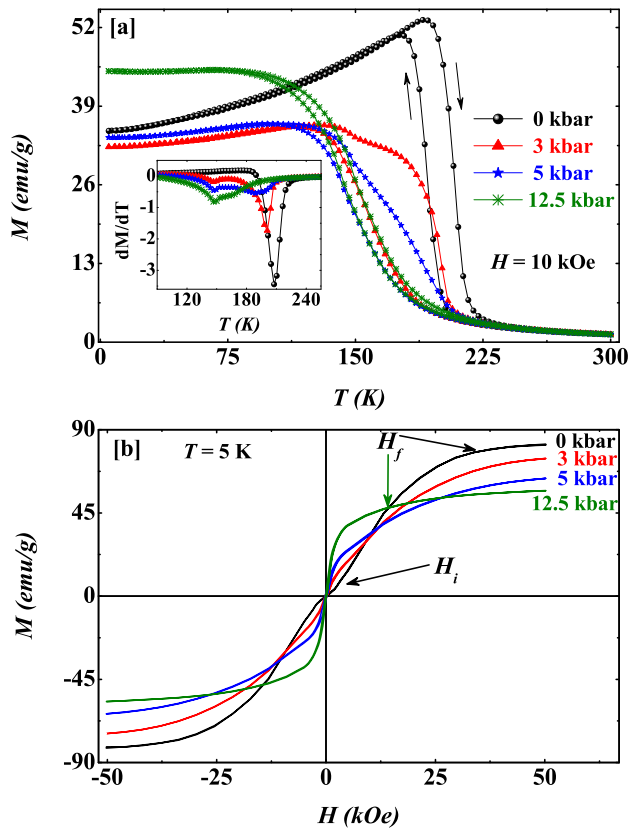}
\caption{(Color online) (a)  Depicts temperature($T$) dependence of field cool (FC) and field cooled heating (FCH) magnetization ($M$) data at 10 kOe of external magnetic field ($H$)  with various applied hydrostatic pressure ($P$). (b) Shows isothermal $M$ as a function of applied $H$ at 5K in presence of different applied $P$.} 
\label{fig3}
\end{figure}

\par
Now, let us examine the effect of external hydrostatic pressure ($P$) on the magnetic and magneto-functional properties of the studied alloy. The main panel of fig.~\ref{fig3}(a) depicts the $T$ variation of FC and FCH magnetization data at 10 kOe of $H$ in presence of different external $P$. Application of external $P$ has a drastic effect on the magnetic and magneto-structural properties of the alloy. MPT becomes broad and is found to be shifted towards lower $T$ side with increasing external $P$. In addition, a closer look reveals that the MPT becomes a two step process with initial increase in $P$ up to 5 kbar. This may be due to the decoupling of magnetic and structural transitions in the alloy. Further increase in $P$ results the merging of these two transitions again. This de-coupling and coupling process is very clearly visible in the $\frac{dM}{dT}$ vs. $T$ data depicted in the inset of fig.~\ref{fig3}(a). This observation is may be due to the faster response of structural transition than the magnetic transition. Initial application of $P$ results a sharp decrease in structural transition temperature and gets saturated around 5 kbar of $P$. On the other hand, magnetic transition temperature decreases slowly with increasing $P$ up to 12.5 kbar. Monotonic decrease in magnetic transition temperature helped to couple these two transitions again. It is also clear from the $M(T)$ data that the low-$T$ martensitic state is more sensitive to the external $P$ than the high-$T$ austenite phase. The magnetic moment in the low-$T$ martensitic state decreases with initial increase in $P$, whilst $M$ starts to increase with further increase in $P$. This anomalous behavior may be related to different FM and AFM phase fractions present in various applied $P$. This anomalous behavior was further investigated by recording isothermal $M(H)$ data at 5 K in presence of different $P$ (see fig.~\ref{fig3}(b)). A number of features are associated with these $M(H)$ isotherms. Firstly, the saturation magnetic moment at 50 kOe of applied $H$ is found to decrease with increasing $P$. About 25 emu/g decrease in $M$ has been observed for the application of 12.5 kbar of external $P$. Closer look at the low field region reveals that the hydrostatic pressure has a very significant effect on the field induced AFM to FM transition present in the martensitic phase of the system. The low-$H$ region of the $M(H)$ data indicates that the starting field $H_i$ of the field induced transition remains almost constant with increasing $P$. On the other hand, the saturation field $H_f$, where this field induced transition finishes, decreases rapidly with increasing $P$. The nature of the $M(H)$ curves suggest that the AFM phase fraction decreases with increasing $P$. Reduced saturation moment with enhancement of $P$ is observed only due to the decrease in the in-plane Mn-Mn direct exchange interaction (by decreasing Mn-Mn intra layer distance) and hence FM strength. Similar decrease in saturation moment at 5 K has also been observed with the increase of chemical pressure (Fe doping in the present case) beyond certain level.~\cite{liu-nc} Up to 16\% of Fe doping, a clear increase in the saturation moment has been observed. Whilst, further increase in Fe concentration results a decrease in saturation moment. The presently studied alloy is a 15\% Fe doped alloy, which is very near to the composition with maximum moment, and shows similar behavior with external hydrostatic pressure as that of the chemical pressure.

\begin{figure}[t]
\centering
\includegraphics[width = 8.5 cm]{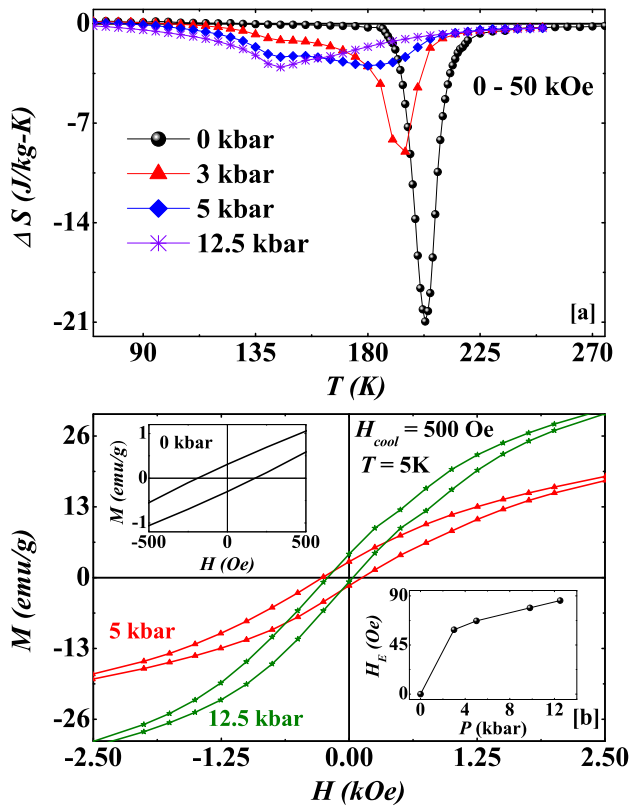}
\caption{(Color online) (a) Shows change in entropy ($\Delta S$) as a function of temperature ($T$) for the applied field changing from 0 to 50 kOe, calculated from the magnetization data at different applied pressure ($P$). (b)Represents the isothermal $M(H)$ data at 5 K after the sample being cooled in $H_{cool}$ = 500 Oe from 300 K at different applied pressure ($P$). The $M(H)$ were recorded by varying the field between $\pm$50 kOe, while only a restricted region between $\pm$ 2.5 kOe are shown for clarity. $M(H)$ isotherm at $P$ = 0 kbar and $H_{cool}$ = 500 Oe is ploted in the upper inset of (b). The lower inset shows the variation of exchange field ($H_E$) as a function of the applied $P$.}   
\label{fig4}
\end{figure}

\par
Tempted by the observation of strong hydrostatic pressure effect on the magnetic properties of the presently studied alloy, we have investigated different magneto-functional properties in presence of different external $P$. The MM$^{\prime}$X alloys are popular for their large magneto-caloric effect (MCE). Change in entropy ($\Delta S$) on application of external $H$ is a measure of MCE, and it is often calculated from the magnetization data using Maxwell's relation,

\begin{equation}
{\Delta}S(0{\rightarrow}H_0)=\int^{H_0}_0\left(\frac{{\partial}M}{{\partial}T}\right)_HdH
\label{maxwell}
\end{equation} 

where ${\Delta}S(0{\rightarrow}H_0)$ denotes the entropy change for the change in $H$ from $0$ to $H_0$.~\cite{gs-mce} It is important to be very much careful while calculating $\Delta S$ around first order phase transition by using Maxwell's relation. Discontinuity of $M$ around the ideal first order phase transition may give rise to infinite value of $\left(\frac{\partial M}{\partial T}\right)_H$ and hence unphysical value of $\Delta S$.~\cite{mcefirst-prb} No such discontinuity has been observed for the present alloy around or below the transition [resulting finite value of $\left(\frac{\partial M}{\partial T}\right)_H$] and hence allows us to use the Maxwell's relation for the present alloy.

\par
For $\Delta S$  calculation, different $M(H)$ isotherms were recorded during heating between 50-300 K with 5 K interval in thermally demagnetized state (data with 1 K interval were recorded around the transition temperature). This specific state was obtained by cooling the sample from room temperature to 5 K and then heated back to the respective temperature of measurements in zero magnetic field. These $M(H)$ data were then convoluted to obtain $M(T)$ data at different constant $H$. Finally, the change in entropy due to the application of $H$ has been calculated (by using equation 1) and depicted in fig.~\ref{fig4}(a). The magnetic entropy change $\Delta S$ is found to be -21 J/kg-K for $H$ changing from 0 to 50 kOe. Negative value of $\Delta S$ indicates that the observed MCE for the present alloy is of conventional type. The above process was repeated for different external $P$ to examine its effect on MCE. Application of external $P$ results decrease in MCE value with a clear shift of the peak position towards lower $T$. About 18 J/kg-K decrease in $\Delta S$ has been observed on application of 12.5 kbar of external $P$. In addition, a clear split in $\Delta S$ vs. $T$ curve has been observed with initial application of $P$ up to 5 kbar. The MCE curve again becomes a single peak structure on further increase in $P$. This behavior also supports the de-coupling and coupling of magnetic and structural transition observed in $M(T)$ data (hence $\frac{dM}{dT}$ vs. $T$ data). For the development of MCE based refrigerator, beside the peak value of $\Delta S$ one has to look for it's refrigeration capacity (RC) also. RC for a sample is defined as RC = $-\int_{T_{cold}}^{T_{hot}} \Delta S(T) dT$, where $T_{hot}$ and $T_{cold}$ are the source and sink temperatures respectively.~\cite{gs-mce1} We have calculated RC for the present alloy in ambient condition as well as in presence of external $P$. In ambient condition RC value is 218 J/kg for $H$ changing from 0 to 50 kOe. On the other hand, in presence of 12.5 kbar of external $P$ RC is found to be 173 J/kg. The RC values observed for the present alloy both in ambient and high pressure conditions are reasonably high and can be used for practical applications.

\par
As the effect of external $P$ is similar to that of chemical doping (chemical pressure) observed from the magnetization data, the AFM and FM interaction present in the alloy may become comparable with increasing $P$ which often gives rise to exchange bias effect (EBE) as observed in a 20\% Fe doped alloy.~\cite{pd-epl} EBE is defined as the shift in the center of the $M(H)$ isotherm while cooled in presence of external $H$. ~\cite{eb-rev1,eb-rev2,pd-epl} EBE is particularly important because of its application in various technological fields, such as permanent magnets, magnetic recording media,  sensors, and read heads.~\cite{eb-rev1,eb-rev2,sc-prb3,sc-jap2,sc-apl1,pd-jmmm} Different magnetic interfaces like, FM/AFM, FM/SG and FM/ferrimagnet based multilayers play the pivotal role towards the observation of EBE.~\cite{eb-rev1,eb-rev2} We have investigated the EBE of the presently studied alloy by recording isothermal $M(H)$ data at 5 K after cooling the sample from 300 K in presence of 500 Oe cooling field ($H_{cool}$) at different $P$ as depicted in the main panel and upper inset of fig.~\ref{fig4}(b) (only $P$ = 0, 5 and 12.5 kbar data are plotted here for clarity). All the $M-H$ isotherms were recorded between $\pm$50 kOe, which is well above the technical saturation field of the sample and rules out the possibility of minor loop effect towards the observation of EBE. However, a restricted range of $\pm$2.5 kOe is shown in fig.~\ref{fig4}(b) for proper visualization. The zero field cooled $M-H$ isotherms are found to be perfectly symmetric with respect to the origin on both horizontal ($H$) and vertical ($M$) axes (see fig.~\ref{fig3}(b)). The field cooled $M(H)$ isotherm remains symmetrical at ambient condition (see upper inset of fig.~\ref{fig4}(b)). This indicates the absence of any EBE in the alloy. On the other hand, a clear shift in the centre of the hysteresis loop, both in $H$ and $M$ axis, has been observed in presence of external $P$. Variation of shift in the field axis $H_E$ as function of external $P$ has been plotted in the lower inset of fig.~\ref{fig4}(b). A monotonic increase in the $H_E$ value has been observed with increasing $P$ value. This pressure induced EBE is in line with the fact that the application of external pressure results decrease in intra layer Mn-Mn FM interaction and makes it comparable with the inter layer Mn-Mn AFM interaction.

\section{Summary and Conclusions}
Our magnetic studies of the Mn$_{0.85}$Fe$_{0.15}$NiGe alloy under different applied hydrostatic pressure (clean perturbation) bring out several fascinating physical and functional properties of the system. Initial application of $P$ results broadening and splitting of magneto-structural transition along with a clear decrease in the transition temperature. Further increase in external $P$ results merging of magnetic and structural transitions. This de-coupling and coupling of magnetic and structural transitions is one of the unique features observed in the presently studied alloy. Unequal response of magnetic and structural transitions towards the external $P$ plays the pivotal role towards this observation. On the other hand, application of $P$ affects Mn-Mn intra-layer direct exchange interaction and hence considerable decrease in the saturation moment at 5 K has been observed with increasing $P$. In addition, a clear decrease in technical saturation field has also been observed with increasing $P$. Apart from these physical properties, external hydrostatic pressure also has a strong effect on the magneto-functional properties of the alloy. The large conventional MCE observed for the present alloy in ambient condition decreases and the peak temperature found to be shifted towards the lower temperature with increasing $P$. The most important observation is the occurrence of EBE in presence of external $P$. The strength of the coexisting AFM and FM interactions in the present alloy becomes comparable with increasing $P$ and plays the pivotal role towards the observation of EBE. The present alloy is a good candidate for memory applications under high pressure.

\section{Acknowledgment}

Authors would like to thank Department of Science and Technology, India for low temperature high magnetic field facilities at UGC-DAE Consortium for Scientific Research, Kolkata Centre. SC would also like to thank Prof. S. Majumdar of IACS, Kolkata for valuable discussions.


%

\end{document}